\documentclass[a4paper,11pt]{article}
\usepackage{pos}
\usepackage{slashed}

\newcommand{\lb}{\left(}
\newcommand{\rb}{\right)}

\newcommand{\al}{\alpha}

\newcommand{\GeV}{{\ensuremath\rm GeV}}

\title{An overview on low mass scalars at future lepton colliders}

\author*[a]{Tania Robens}

\affiliation[a]{Rudjer Boskovic Institute,\\
  Bijenicka cesta 54, 10000 Zagreb, Croatia}

\emailAdd{trobens@irb.hr}

\abstract{Although many suggestions for BSM searches at future colliders exist, most of them concentrate on additional scalars that have masses higher than the current SM scalar mass. I will give a short overview on the current status of models and searches for scalars with masses below this. This work is mainly based on \cite{Robens:2022erq,Robens:2022zgk}.\\
RBI-ThPhys-2022-47}

\FullConference{%
  41st International Conference on High Energy physics - ICHEP2022\\
  6-13 July, 2022\\
  Bologna, Italy
}

\begin{document}
\maketitle

\section{Processes at Higgs factories}

\begin{center}
\begin{figure}[htb!]
\begin{center}
\includegraphics[width=0.55\textwidth]{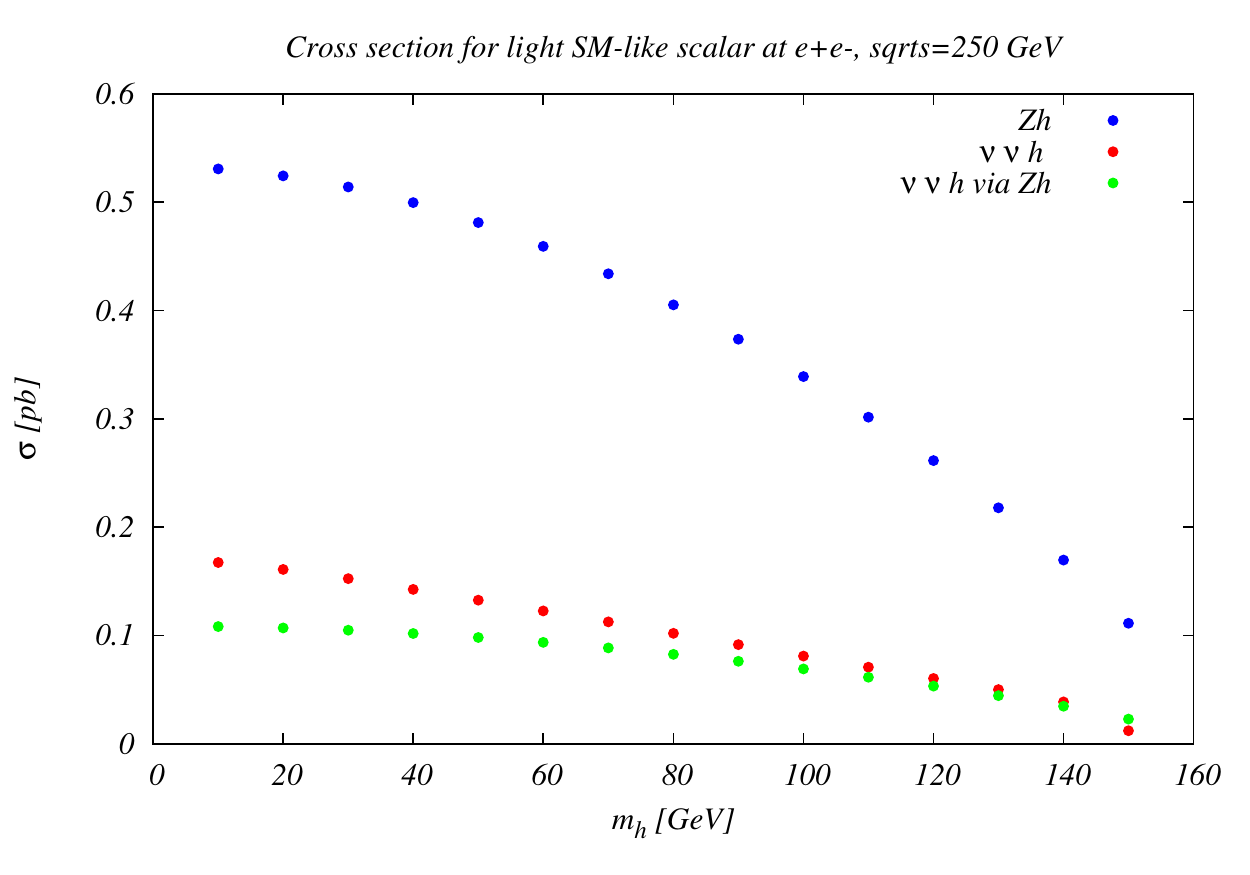}
\caption{\label{fig:prod250} Leading-order production cross sections for $Z\,h$ and $h\,\nu_\ell\,\bar{\nu}_\ell$ production at an $e^+\,e^-$ collider with a com energy of 250 \GeV~  using Madgraph5 for an SM-like scalar h. Shown is also the contribution of $Z\,h$ to $\nu_\ell\,\bar{\nu}_\ell\,h$ using a factorized approach for the Z decay. Update of plot presented in \cite{Robens:2022zgk}, extended to higher mass range.}
\end{center}
\end{figure}
\end{center}

For production at lepton colliders, Higgs strahlung is the dominant production mode at the center-of-mass (com) energies of Higgs factories \cite{Abramowicz:2016zbo}. Leading-order predictions for $Zh$ production at $e^+e^-$ colliders for low mass scalars which are Standard Model (SM)-like, using Madgraph5 \cite{Alwall:2011uj}, are shown in figure \ref{fig:prod250} for a center-of-mass energy of 250 \GeV. We also display the VBF-type production of $e^+e^-\,\rightarrow\,h\,\nu_\ell\,\bar{\nu}_\ell$. Note that the latter contains contributions from $Z\,h$ production, where $Z\,\rightarrow\,\nu_\ell\,\bar{\nu}_\ell$.

\section{Projections for additional searches}
\begin{center}
\begin{figure}[htb!]
\begin{center}
\includegraphics[width=0.4\textwidth, angle=-90]{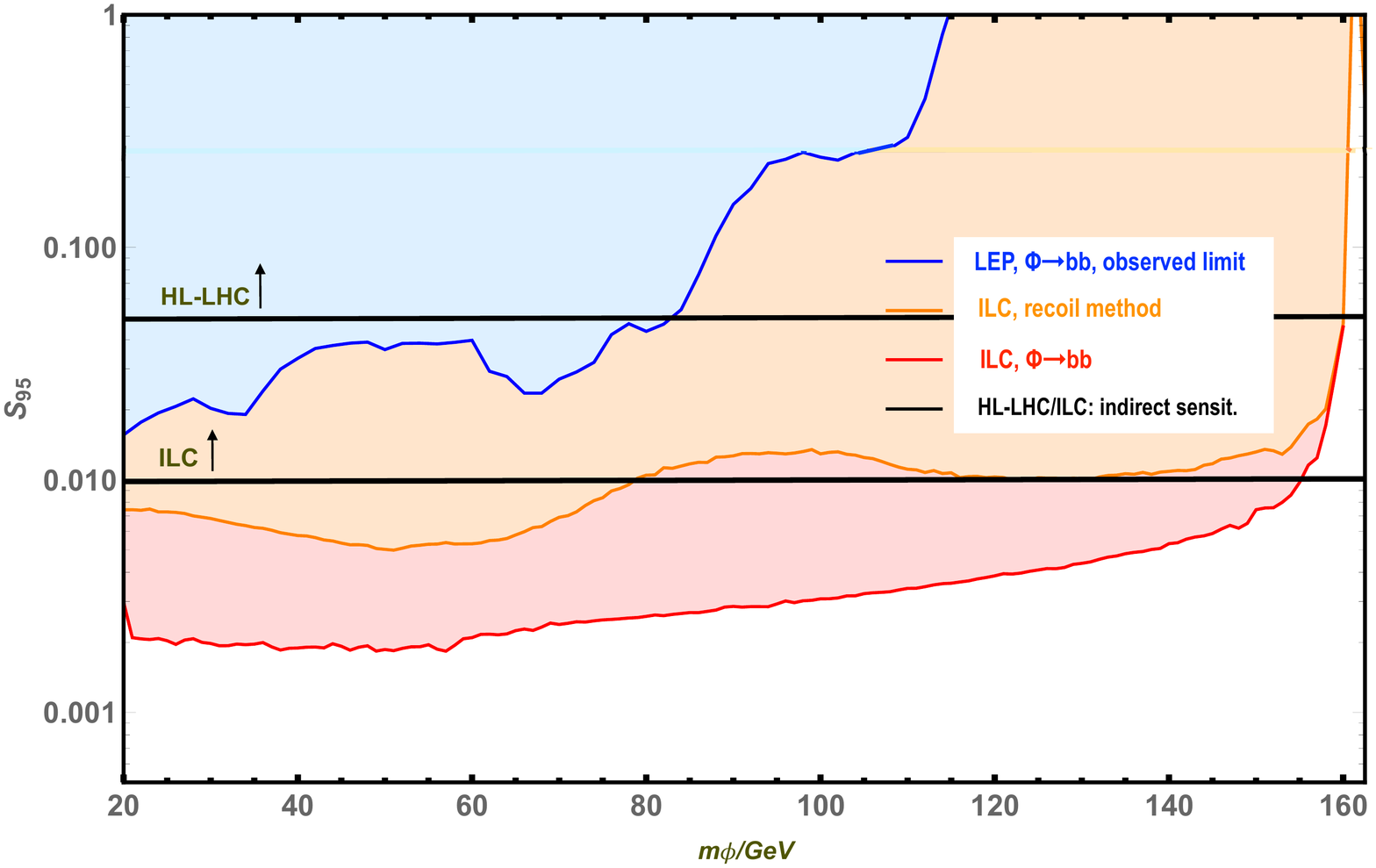}
\caption{\label{fig:lepgea} Sensitivity predictions for an ILC with a com energy of 250 \GeV~ from \cite{Drechsel:2018mgd}. See text for details.}
\end{center}
\end{figure}
\end{center}

For the production mechanism discussed above, in principle two different analysis methods exist, which either use the pure $Z$ recoil ("recoil method") or take the light scalar decay into $b\,\bar{b}$ into account. We here point to \cite{Drechsel:2018mgd}, that investigates the sensitivity of the ILC for low-mass scalars in $Z\,h$ production and compares the reach of these two methods at $95\,\%$ CL limit for agreement with a background only hypothesis, which can directly be translated into an upper bound on rescaling. 
The results are shown in figure \ref{fig:lepgea}.

\section{Parameter space for some sample models}

\begin{center}
\begin{figure}[htb!]
\begin{center}
\includegraphics[width=0.48\textwidth]{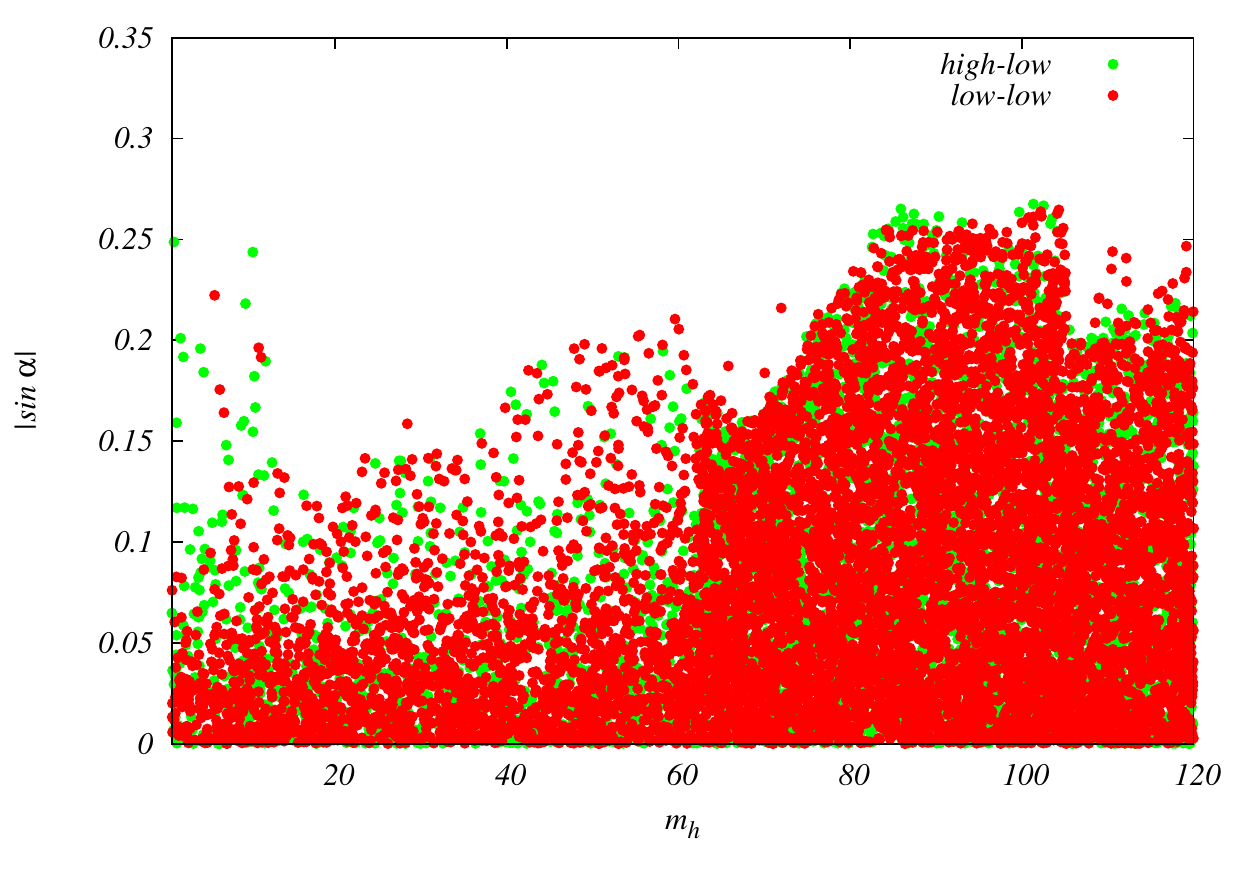}
\includegraphics[width=0.48\textwidth]{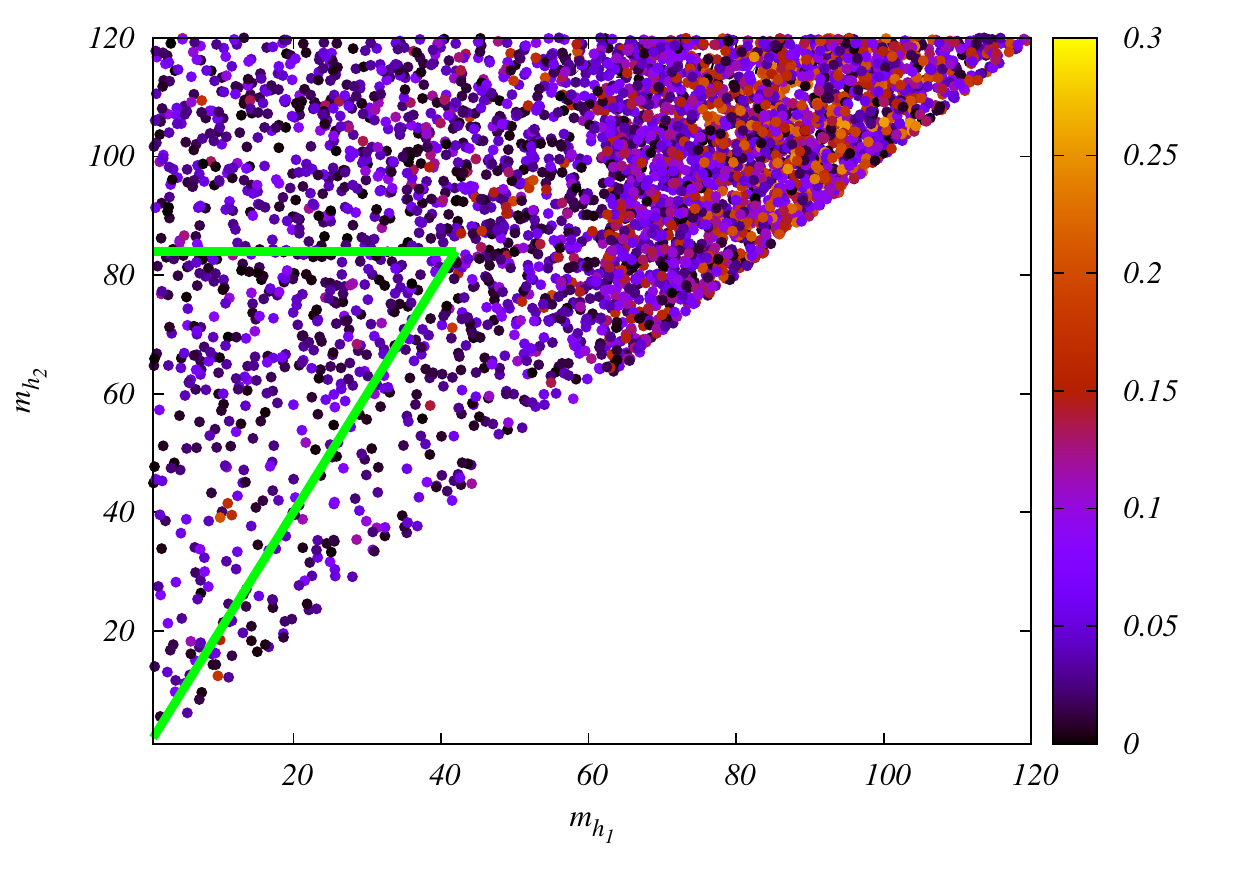}
\caption{\label{fig:trsm} Available parameter space in the TRSM, with one (high-low) or two (low-low) masses lighter than 125 \GeV. {\sl Left}: light scalar mass and mixing angle, with $\sin\al\,=\,0$ corresponding to complete decoupling. {\sl Right:} available parameter space in the $\lb m_{h_1},\,m_{h_2}\rb$ plane, with color coding denoting the rescaling parameter $\sin\al$ for the lighter scalar $h_1$. Within the green triangle, $h_{125}\,\rightarrow\,h_2 h_1\,\rightarrow\,h_1\,h_1\,h_1$ decays are kinematically allowed. Taken from \cite{Robens:2022zgk}.}
\end{center}
\end{figure}
\end{center}

\begin{center}
\begin{figure}[htb!]
\begin{center}
\includegraphics[width=0.48\textwidth]{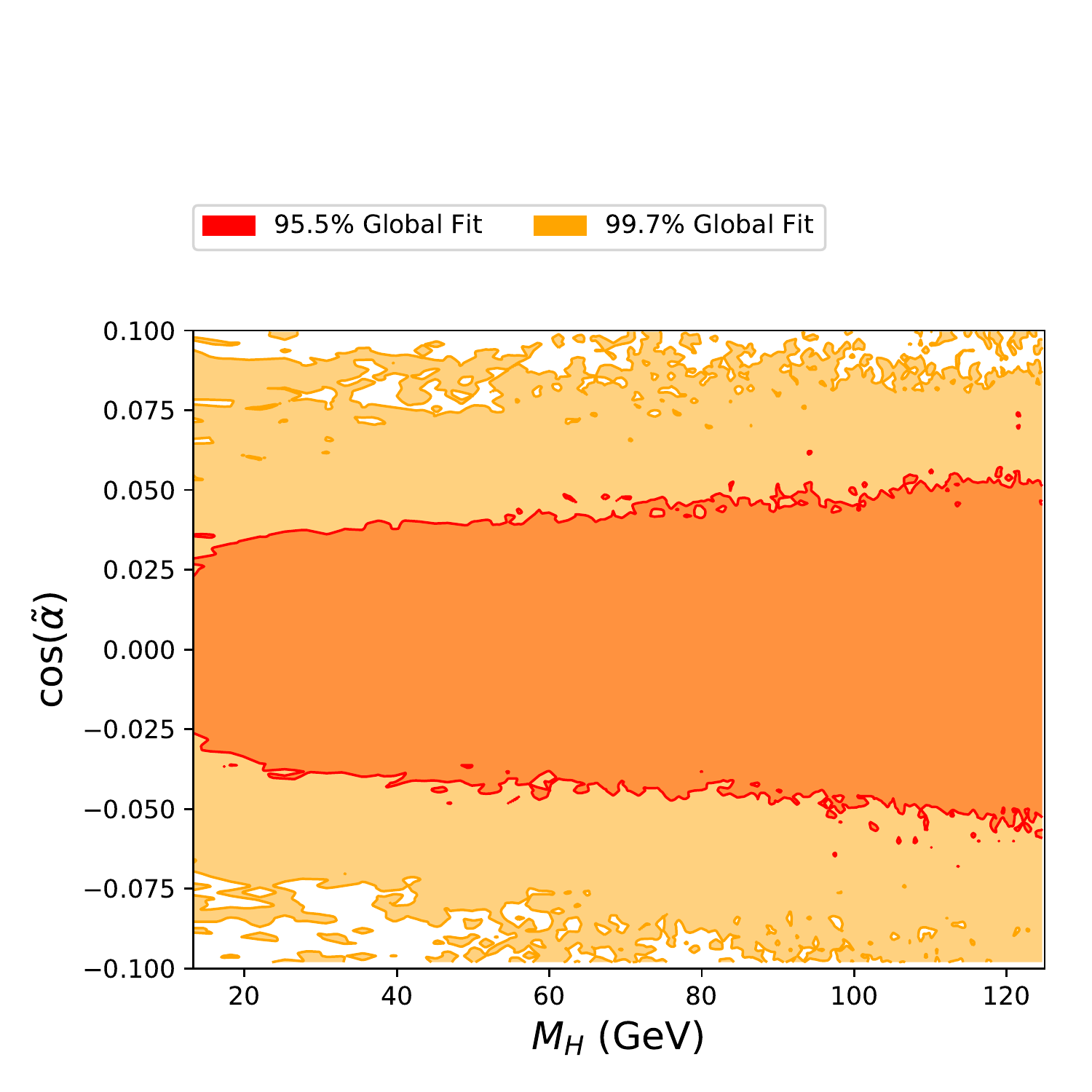} 
\includegraphics[width=0.48\textwidth]{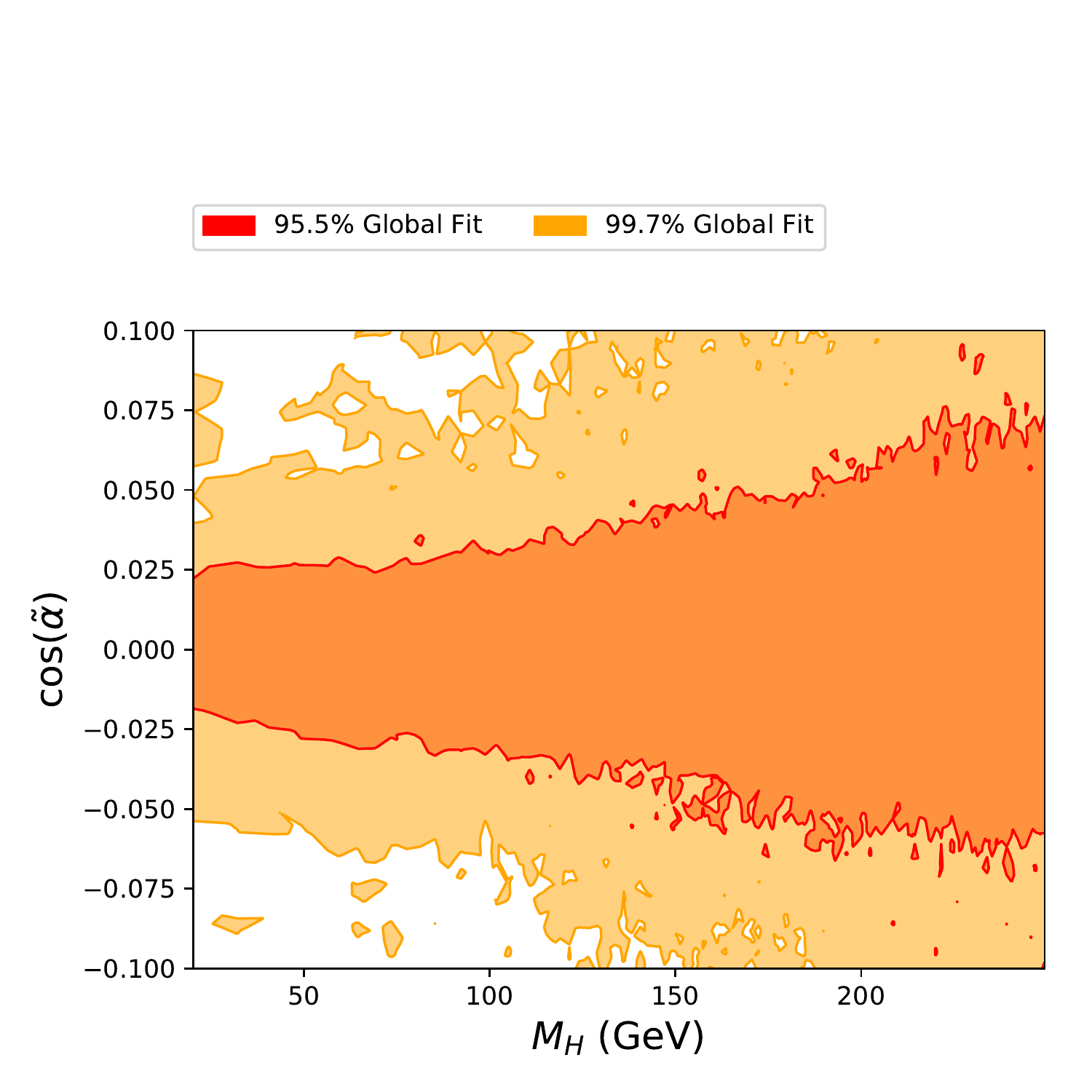}
\caption{\label{fig:victor} Mixing angle and masses of different additional scalars in the aligned 2HDM, from the scan presented in \cite{Eberhardt:2020dat}. For all additional scalars, regions exists where masses are $\lesssim\,125\,\GeV$, with absolute values of mixing angles such that $|\cos\lb\tilde{\al}\rb|\lesssim\,0.1$. Taken from \cite{Robens:2022zgk}.}
\end{center}
\end{figure}
\end{center}

Another important question is to investigate which models still leave room for low mass scalars taking all current constraints into account. For this, we show an example of the allowed parameter space in a model with two additional singlets \cite{Robens:2019kga}. This model contains three CP-even neutral scalars that relate the gauge and mass eigenstates $h_{1,2,3}$ via mixing. 
A detailed discussion of the model including theoretical and experimental constraints can be found in \cite{Robens:2019kga,Robens:2022nnw}. In figure \ref{fig:trsm}, we display two cases where either one (high-low) or two (low-low) scalar masses are smaller than $125\,\GeV$. On the y-axis, the respective mixing angle is shown, where complete decoupling would be designated by $\sin\al\,=\,0$.
It is also of interest to investigate different extensions, as e.g. two Higgs doublet models, where the SM scalar sector is augmented by a second doublet. In the so-called flavour-aligned scenario  \cite{Pich:2009sp,Pich:2010ic}, the authors perform a scan including bounds from theory, experimental searches and constraints, as e.g. electroweak observables, as well as B-physics. Here, the angle $\tilde{\al}$ parametrizes the rescaling with respect to the Standard Model couplings to gauge bosons, with $\cos\tilde{\al}\,=\,0$ designating the SM decoupling. The limits on the absolute value of the cosine of rescaling angle vary between 0.05 and 0.25 \cite{ATLAS-CONF-2021-053}. In figure \ref{fig:victor}, we display this angle vs the different scalar masses, reproduced from \cite{Eberhardt:2020dat}. 
We see that all regions for masses $\lesssim\,125\, \GeV$ can be populated, with absolute value of mixing angle ranges $|\cos\lb\tilde{\al}\rb|\lesssim\,0.1$.
\section{Conclusions}
I very briefly discussed some aspects of searches for low mass scalars at Higgs factories, including models that allo for such low mass states, and provided references for further reading.

\begin{thebibliography}{10}

\bibitem{Robens:2022erq}
T.~Robens,
\newblock {A short overview on low mass scalars at future lepton colliders -
  Snowmass White Paper},
\newblock in {\em {2022 Snowmass Summer Study}}, 2022, 2203.08210.

\bibitem{Robens:2022zgk}
T.~Robens,
\newblock Universe {\bf 8}, 286 (2022), 2205.09687.

\bibitem{Abramowicz:2016zbo}
H.~Abramowicz {\em et~al.},
\newblock Eur. Phys. J. C {\bf 77}, 475 (2017), 1608.07538.

\bibitem{Alwall:2011uj}
J.~Alwall, M.~Herquet, F.~Maltoni, O.~Mattelaer, and T.~Stelzer,
\newblock JHEP {\bf 06}, 128 (2011), 1106.0522.

\bibitem{Drechsel:2018mgd}
P.~Drechsel, G.~Moortgat-Pick, and G.~Weiglein,
\newblock Eur. Phys. J. C {\bf 80}, 922 (2020), 1801.09662.

\bibitem{Robens:2019kga}
T.~Robens, T.~Stefaniak, and J.~Wittbrodt,
\newblock Eur. Phys. J. C {\bf 80}, 151 (2020), 1908.08554.

\bibitem{Robens:2022nnw}
T.~Robens,
\newblock {Two-Real-Singlet-Model Benchmark Planes},
\newblock 2022, 2209.10996.

\bibitem{Pich:2009sp}
A.~Pich and P.~Tuzon,
\newblock Phys. Rev. D {\bf 80}, 091702 (2009), 0908.1554.

\bibitem{Pich:2010ic}
A.~Pich,
\newblock Nucl. Phys. B Proc. Suppl. {\bf 209}, 182 (2010), 1010.5217.

\bibitem{ATLAS-CONF-2021-053}
CERN Report No.,
\newblock , 2021 (unpublished),
\newblock ATLAS-CONF-2021-053.

\bibitem{Eberhardt:2020dat}
O.~Eberhardt, A.~P.~n. Mart\'\i{}nez, and A.~Pich,
\newblock JHEP {\bf 05}, 005 (2021), 2012.09200.

\end{thebibliography}

\end{document}